\documentclass[aps,prd,preprint,groupedaddress,nofootinbib]{revtex4-1}

\usepackage{graphicx}
\usepackage{epsfig}
\usepackage{bm}
\usepackage{amsfonts}
\usepackage{amsmath}
\usepackage{amssymb}
\usepackage{subfigure}

\providecommand{\tabularnewline}{\\}

\begin{document}

\title{Mechanism of primordial black holes production and secondary gravitational waves in $\alpha$-attractor Galileon inflationary scenario}

\author{
Zeinab Teimoori$^{1,3}$\footnote{zteimoori16@gmail.com},
Kazem Rezazadeh$^{2}$\footnote{kazem.rezazadeh@ipm.ir},
Mariwan Ahmed Rasheed $^{3,4}$\footnote{mariwan.rasheed@uhd.edu.iq},
Kayoomars Karami$^{1}$\footnote{kkarami@uok.ac.ir}
}

\affiliation{
$^{1}$\small{Department of Physics, University of Kurdistan, Pasdaran Street, P.O. Box 66177-15175, Sanandaj, Iran}\\
$^{2}$\small{School of Physics, Institute for Research in Fundamental Sciences (IPM),
P.O. Box 19395-5531, Tehran, Iran}\\
$^{3}$\small{Development Center for Research and Training (DCRT), University of Human Development, Sulaimani, Kurdistan Region, Iraq}\\
$^{4}$\small{Department of Computer Science, College of Science and Technology, University of Human Development, Sulaymaniyah, Kurdistan Region of Iraq}
}


\begin{abstract}

We study the process of the Primordial Black Holes (PBHs) production in the novel framework, namely $\alpha$-attractor Galileon inflation (G-inflation) model. In our framework, we take the Galileon function as $G(\phi)=G_{I}(\phi)\left(1+G_{II}(\phi)\right)$, where the part $G_{I}(\phi)$ is motivated from the $\alpha$-attractor inflationary scenario in its original non-canonical frame, and it ensures for the model to be consistent with the Planck 2018 observations at the CMB scales. The part $G_{II}(\phi)$ is invoked to enhance the curvature perturbations at some smaller scales which in turn gives rise to PBHs formation. By fine-tuning of the model parameters, we find three parameter sets which successfully produce a sufficiently large peak in the curvature power spectrum. We show that these parameter sets produce PBHs with masses ${\cal O}(10)M_\odot$, ${\cal O}(10^{-5})M_\odot$, and ${\cal O}(10^{-13})M_\odot$ which can explain the LIGO events, the ultrashort-timescale microlensing events in OGLE data, and around $0.98\%$ of the current Dark Matter (DM) content of the universe, respectively. Additionally, we study the secondary Gravitational Waves (GWs) in our setup and show that our model anticipates the peak of their present fractional energy density as $\Omega_{\rm GW_0} \sim 10^{-8}$ for all the three parameter sets, but at different frequencies. These predictions can be located well inside the sensitivity region of some GWs detectors, and therefore the compatibility of our model can be assessed in light of the future data. We further estimate the tilts of the included GWs spectrum in the different ranges of frequency, and confirm that spectrum follows the power-law relation $\Omega_{\rm GW_0}\sim f^{n}$ in those frequency bands.

\end{abstract}


\maketitle

\newpage

\section{Introduction}
\label{sec:intro}

The gravitational collapse of sufficiently large amplitude curvature perturbations generated during the inflationary period, could lead to formation of the primordial black holes (PBHs), at the horizon re-entry in the radiation dominated era. This idea was first suggested by Zel'dovich and Novikov in 1966, and then by Hawking and Carr in the early 1970's \cite{zel1967hypothesis, Hawking:1971ei, Carr:1974nx}. Although PBHs have been studied extensively over the decades by physicists and astronomers, the first detection of Gravitational Wave (GW), GW150914, from merging of two black holes with mass $\sim 30 M_\odot$ ($M_\odot$ is the solar mass) by the LIGO-Virgo Collaboration \cite{Abbott:2016blz, Abbott:2016nmj}, has attracted more attention to the physics of PBHs and the possibility that they could account for all or a fraction of the Dark Matter (DM) of the universe \cite{Bird:2016dcv, Sasaki:2016jop, Blinnikov:2016bxu, Carr:2016drx, Clesse:2016vqa, Carr:2017jsz, mroz2017no, Garcia-Bellido:2017mdw, Germani:2017bcs, Motohashi:2017kbs, Clesse:2017bsw, Ezquiaga:2017fvi, Gong:2017qlj, ballesteros2018primordial, Zumalacarregui:2017qqd, Katz:2018zrn, Cai:2018tuh, Niikura:2017zjd, Ballesteros:2018wlw, Kamenshchik:2018sig, Niikura:2019kqi, Laha:2019ssq, Chen:2019zza, Fu:2019ttf, Dalianis:2018frf, Dasgupta:2019cae, Mishra:2019pzq, Dalianis:2019vit, Mahbub:2019uhl, Ashoorioon:2019xqc, Ashoorioon:2020hln, Solbi:2021rse,Solbi:2021wbo, Liu:2021rgq,Teimoori2021,Saito:2008em,Byrnes:2012yx, Young:2013oia,Young:2013oia,Tada:2015noa,Young:2015kda,Young:2015cyn,
Garcia-Bellido:2017aan,Franciolini:2018vbk,ezquiaga2018quantum,Passaglia:2018ixg,Atal:2018neu,Belotsky:2018wph,DeLuca:2019qsy,Yoo:2019pma,Ezquiaga:2019ftu,Serpico:2020ehh, Braglia:2020taf,Figueroa:2020jkf,
Garcia-Bellido:1996mdl,Sasaki:2006kq,Kawasaki:2006zv,Kawasaki:2006zv,Kawaguchi:2007fz,Frampton:2010sw,Lyth:2011kj,Bugaev:2011wy,
Kohri:2012yw,Kawasaki:2012wr,Linde:2012bt,Bugaev:2013vba,Clesse:2015wea,Garcia-Bellido:2016dkw,Kawasaki:2016pql,Inomata:2016rbd,Domcke:2017fix,Braglia:2020eai}.
The recent detection of the ultra-short timescale microlensing events by the OGLE collaboration has provided an allowed area for PBHs formation \cite{mroz2017no, Niikura:2019kqi}. The favored area from the OGLE data indicate that the PBHs abundance can reach $\mathcal{O}\left(10^{-2}\right)$ at the mass scale $\mathcal{O}\left(10^{-5}\right)M_\odot$ \cite{mroz2017no, Niikura:2019kqi}. PBHs can also constitute a considerable part of DM in the mass ranges $10^{-16}-10^{-14}M_\odot$ and $10^{-13}-10^{-11}M_\odot$ \cite{Niikura:2017zjd, Katz:2018zrn, Laha:2019ssq, Dasgupta:2019cae,Carr:2009jm, Graham:2015apa, Pi:2017gih}.

In order to have PBHs formation, a sufficiently large enhancement should be happened in the power spectrum of the curvature perturbations, so that the amplitude of the spectrum could reach $\mathcal{P}_s\sim{\cal O}(10^{-2})$ at some scales. On the other hand, the CMB observations constrain the amplitude of the scalar perturbation at the pivot scale $k_{*}=0.05 {\rm{Mpc}}^{-1}$ as $\mathcal{P}_s\sim{\cal O}(10^{-9})$ \cite{akrami2020planck}. Various mechanisms have already been proposed in the literature to generate large peaks in the power spectrum leading finally to PBHs in some scales during inflation. They are characterized by different statistics of the curvature perturbations. Because, the statistical properties of the primordial perturbations can affect the PBHs abundance \cite{Saito:2008em,Byrnes:2012yx, Young:2013oia,Young:2013oia,Tada:2015noa,Young:2015kda,Young:2015cyn,
Garcia-Bellido:2017aan,Franciolini:2018vbk,ezquiaga2018quantum,Passaglia:2018ixg,Atal:2018neu,Belotsky:2018wph,DeLuca:2019qsy,Yoo:2019pma,Ezquiaga:2019ftu,Serpico:2020ehh, Braglia:2020taf,Figueroa:2020jkf}. The production of PBHs with the intensive enhancement of the curvature power spectrum from a Gaussian distribution, in single-field inflation scenarios, has been well studied in \cite{Bird:2016dcv, Sasaki:2016jop, Blinnikov:2016bxu, Carr:2016drx, Clesse:2016vqa, Carr:2017jsz, mroz2017no, Garcia-Bellido:2017mdw, Germani:2017bcs, Motohashi:2017kbs, Clesse:2017bsw, Ezquiaga:2017fvi, Gong:2017qlj, ballesteros2018primordial, Zumalacarregui:2017qqd, Katz:2018zrn, Cai:2018tuh, Niikura:2017zjd, Ballesteros:2018wlw, Kamenshchik:2018sig, Niikura:2019kqi, Laha:2019ssq, Chen:2019zza, Fu:2019ttf, Dalianis:2018frf, Dasgupta:2019cae, Mishra:2019pzq, Dalianis:2019vit, Mahbub:2019uhl, Ashoorioon:2019xqc, Ashoorioon:2020hln, Solbi:2021rse, Solbi:2021wbo, Liu:2021rgq,Teimoori2021}. For instance, Cai \textit{et al.} \cite{Cai:2018tuh, Chen:2019zza} have succeeded in doing so, by the parametric resonance arising from the oscillating sound speed squared. In \cite{Fu:2019ttf, Dalianis:2019vit, Teimoori2021}, this goal can be achieved by mechanism of the gravitationally enhanced friction arising from the nonminimal derivative coupling scenario. Using the inflationary potentials with inflection points is another mechanism leading to such an enhancement \cite{Garcia-Bellido:2017mdw, Germani:2017bcs, Motohashi:2017kbs, Ezquiaga:2017fvi, Gong:2017qlj, ballesteros2018primordial, Dalianis:2018frf, Mahbub:2019uhl}. In this scenario, the inflaton field experiences a very flat potential in the inflection point region of the potential. Such a flat region results in a so-called ultra slow-roll phase in which the inflaton velocity decreases with a faster rate than the slow-roll phase and inflaton has a friction dominated phase. Consequently, the curvature perturbation grows rapidly due to the great decrease of the Hubble slow-roll parameter \cite{Biagetti:2021eep}.

Another assumption is that the scalar perturbations obey the non-Gaussian statistics \cite{Saito:2008em,Byrnes:2012yx, Young:2013oia,Young:2013oia,Tada:2015noa,Young:2015kda,Young:2015cyn,
Garcia-Bellido:2017aan,Franciolini:2018vbk,ezquiaga2018quantum,Passaglia:2018ixg,Atal:2018neu,Belotsky:2018wph,DeLuca:2019qsy,Yoo:2019pma,Ezquiaga:2019ftu,Serpico:2020ehh, Braglia:2020taf,Figueroa:2020jkf}.
For instance, the authors of \cite{Garcia-Bellido:2017aan} have discussed the impact of non-Gaussianity in determining the PBHs abundance. They have concluded that the non-Gaussian perturbations can produce same abundance of PBH from a smaller amplitude of scalar perturbations than the Gaussian case. It is important to stress that due the effect of non-Gaussianity on the PBH abundance, PBHs can be used as a tool to investigate the amplitude and non-Gaussianity of the primordial perturbations on small scales.

PBHs can also be produced in multiple-field inflationary models \cite{Young:2013oia,
Garcia-Bellido:1996mdl,Sasaki:2006kq,Kawasaki:2006zv,Kawaguchi:2007fz,Frampton:2010sw,Lyth:2011kj,Bugaev:2011wy,
Kohri:2012yw,Kawasaki:2012wr,Linde:2012bt,Bugaev:2013vba,Clesse:2015wea,Garcia-Bellido:2016dkw,Kawasaki:2016pql,Inomata:2016rbd,Domcke:2017fix,Braglia:2020eai}. In these models, sufficiently large peaks to PBHs formation can be generated with both curvature (adiabatic) \cite{Polarski:1992dq,Starobinsky:1994mh} and isocurvature   \cite{Starobinsky:1994mh,Kofman:1986wm,Kofman:1988xg,Polarski:1994rz,Starobinsky:2001xq} perturbations. For instance, the authors of \cite{Clesse:2015wea}, have succeeded within a two-field inflation to produce a large peak in the power spectrum of curvature perturbations leading finally to PBHs, in the context of the more traditional hybrid inflation. Note that multi-field models of inflation can produce strong non-Gaussianity. The curvaton model is a well motivated example \cite{Young:2013oia,Sasaki:2006kq}.
In this work, assuming Gaussian statistics of the curvature perturbations in the single field inflationary scenario, we propose a novel mechanism to achieve the ultra slow-roll phase. We focus on the Galileon scalar field theory with the Galileon term $G_3(\phi)\Box\phi$ in which $G_3(\phi)$ is a general function in terms of the Galileon field $\phi$. The most important motivation for choosing the Galileon scalar field theory is that, the field equations driven from this theory include derivatives only up to second order \cite{Deffayet:2009wt, Deffayet:2009mn, Kobayashi:2010cm, kobayashi2011generalized, Charmousis:2011bf}. According to the theorem of Ostrogradsky \cite{Ostrogradsky:1850}, higher-derivative theories have additional degrees of freedom and usually suffer from serious problems, such as negative energies and related instabilities. Note that the Lagrangian of the Galileon scalar field theory which was first introduced by Nicolis in \cite{Nicolis:2008in}, inspired by Dvali-Gabadadze-Porrati (DGP) model \cite{Dvali:2000hr}, is equivalent to the so-called Horndeski theory \cite{Horndeski:1974wa, kobayashi2011generalized, Charmousis:2011bf}. For more study about the Galileon scalar field theory, see e.g.  \cite{Deffayet:2011gz, Kamada:2010qe, Kobayashi:2011pc, burrage2011galileon, ohashi2012potential, Teimoori:2017kob}.

Recently, the formation of PBHs has been studied in the context of G-inflation by some authors \cite{Lin:2020goi, Yi:2020cut, Gao:2020tsa, Gao:2021vxb,Solbi:2021rse,Solbi:2021wbo}. In the present work, we investigate this subject, and in our work, we go a step further and take the functional form of $G_3(\phi)$ in such a way that our Galileon model is transformed at significant period of time domain of interest during inflation to a class of superconformal inflationary models called cosmological $\alpha$-attractors \cite{Kallosh:2013yoa, Kallosh:2014rga, Kallosh:2014laa, Kallosh:2015lwa, Linde:2015uga, Galante:2014ifa, Carrasco:2015rva, Carrasco:2015pla, Roest:2015qya, scalisi2015cosmological, Eshaghi:2016kne}. The $\alpha$-attractors are described in terms of a free parameter $\alpha$ that is inversely proportional to the curvature of the inflaton K\"{a}hler manifold \cite{Kallosh:2013yoa, Galante:2014ifa}. These class of models have a universal attractor behavior in the Einstein frame. In the limit of the large $e$-fold number $N$ and small $\alpha$, they yield the same predictions for the inflationary observables $n_s$ and $r$ as $n_s=1-2/N$ and $r = 12\alpha/N^2$, which are in a excellent agreement with the recent observational data \cite{akrami2020planck}. For $\alpha=1$, these quantities correspond to the obtained results in the Starobinsky model \cite{Starobinsky:1980te} and Higgs inflation scenario \cite{Bezrukov:2007ep}. Therefore, we can hope that the inflationary observables $n_s$ and $r$, more likely to be consistent with the latest observational constraints at around the sound horizon exit. Accordingly, in this new framework that we label it as $\alpha$-attractor Galileon inflation model, we examine the possibility of PBHs production with masses around ${\cal O}(10)M_\odot$, ${\cal O}(10^{-5})M_\odot$, and ${\cal O}(10^{-13})M_\odot$.

The enhancement of primordial curvature perturbation that leads to the production of PBHs, may also induce the secondary gravitational waves (GWs) \cite{Matarrese:1997ay, Mollerach:2003nq, Ananda:2006af, Baumann:2007zm,  Saito:2008jc, Saito:2009jt, Bugaev:2009zh, Bugaev:2010bb, Alabidi:2012ex,Nakama:2016gzw, Inomata:2016rbd,Garcia-Bellido:2017aan, Peirone:2017vcq, Cheng:2018yyr}. After the horizon re-entry, the overdense regions may collapse and consequently generate large metric perturbations. In the second-order, the scalar and tensor perturbations can be coupled to each other, and hence the scalar metric perturbations, through the second-order effect, can lead to the propagation of the stochastic GWs among the cosmological background \cite{Cai:2018dig, Cai:2019amo, Cai:2019jah, Bartolo:2018evs, Bartolo:2018rku, Wang:2019kaf}. Thus, the detection of such a signal for the induced GWs can be employed as a novel approach to probe the PBHs. Besides, the detection of both PBHs and secondary GWs can be regarded to constrain the substantial amplification of the spectrum of the primordial curvature perturbations at some scales during inflation. So, in this way, we can acquire valuable information about the physics of the early universe.
So far, various scenarios have been proposed to generate secondary GWs from inflation. For instance, in \cite{Saito:2008jc, Bartolo:2018evs, Cai:2018dig}, the authors have used a phenomenological delta function to create the required enhancement of the scalar power spectrum. Such an enhancement can also be supplied by using of a broken power-law \cite{Lu:2019sti} or Gaussian power spectrum \cite{Lu:2019sti, Namba:2015gja, Garcia-Bellido:2017aan, Ozsoy:2020ccy, Ozsoy:2020kat}. For other mechanisms for induction of secondary GWs in the context of inflationary cosmology, one can refer to \cite{Stewart:1996ey, Drees:2011hb, Datta:2019euh, Kasuya:2009up, Kawasaki:2012wr, Easther:2006vd, Antusch:2016con, Liu:2017hua, Kawasaki:2006zv, Kawasaki:2016pql}.
Note that the amplitude of induced GWs depends on both the amplitude of scalar perturbations and their statistics \cite{Garcia-Bellido:2017aan}. For instance, in \cite{Nakama:2016gzw}, the authors assuming the non-Gaussian statistics have showed that for an equal abundance of PBH, the power of the induced GWs is much smaller than the Gaussian case.

In this paper, under the assumption that the probability distribution function of curvature perturbations is Gaussian, we investigate the secondary GWs in the setup of $\alpha$-attractor G-inflation. In our work, we calculate the present fractional energy density of these GWs and compare our findings with the sensitivity regions of different GWs detectors. We further estimate the slope of the secondary GWs spectrum in different domains of the frequency band. This is another feature of the secondary GWs that can be probed by using of the upcoming data, and also can be used to discriminate between the different PBHs scenarios from inflation.

The paper is structured as follows. In Sec. \ref{sec:PBHAbundance}, we review the basics of the PBH formation. In Sec. \ref{sec:review}, we present a brief review of G-inflation and derive all the necessary equations describing our model. In Sec. \ref{sec:PBHGalileon}, we discuss the possibility of PBH formation in the framework of the $\alpha$-attractor G-inflation scenario. We investigate the secondary GWs in our framework in Sec. \ref{sec:sgws}. Finally, Sec. \ref{sec:con} is devoted to our concluding remarks.


\section{Abundance of Primordial Black Holes}\label{sec:PBHAbundance}

In this section, we review briefly the basic formulas for PBHs formation from inflation. As mentioned in Sec. \ref{sec:intro}, the curvature perturbations generated during inflation may collapse and form PBHs after horizon re-entry in the radiation dominated epoch, if they have sufficiently large amplitude. The mass of formed PBHs, $M(k)$, is related to the horizon mass, $M_{\rm hor}$, by \cite{Gong:2017qlj}
\begin{equation}\label{PBHmass}
M(k)=\gamma M_{\rm hor}=3.68\left(\frac{\gamma}{0.2}\right)\left(\frac{g_*}{10.75}\right)^{-1/6}\left(\frac{k}{10^{6}\,\,\rm{Mpc^{-1}}}\right)^{-2}M_{\odot},
\end{equation}
where the efficiency factor $\gamma$ depends on the details of the gravitational collapse and usually is assumed to be $\gamma \simeq 0.2 $ \cite{carr1975primordial}, and $g_*=107.5$ is the effective degrees of freedom for the energy density.

In the Press-Schechter formalism \cite{Press:1973iz}, assuming the Gaussian statistics of the curvature perturbations, the production rate of PBHs with mass $M(k)$ is given by \cite{Young:2014ana, ozsoy2018mechanisms, Tada:2019amh}
\begin{equation}\label{beta}
\beta(M)=\sqrt{\frac{2}{\pi}} \frac{\sigma(M)}{\delta_c}\exp{\left(-\frac{\delta_c^{2}}{2\sigma^2(M)}\right)},
\end{equation}
where $\delta_c$ is the threshold of the density perturbation for the PBHs formation, and in this paper, we adopt its value to be $0.4$ \cite{Musco:2012au, Harada:2013epa, Escriva:2019phb}. The quantity $\sigma^2(k)$ denotes the coarse-grained variance of the density contrast smoothed on a scale $k$ that is given by \cite{Young:2014ana, ozsoy2018mechanisms}
\begin{equation}\label{sigma2}
\sigma^2(k)=\int{\frac{dq}{q}}\,W^2(q/k)\frac{16}{81}(q/k)^4 {\cal P}_s(q),
\end{equation}
where ${\cal P}_s$ represents the power spectrum of the curvature perturbations and $W(x)$ is the window function. There are several different window functions that their effect on the PBHs abundance has been studied in the literature \cite{Young:2019osy,Ando:2018qdb}. In our analysis, we take a popular and appropriate Gaussian window function as $W(x)=\exp{(-x^2/2)}$.

The current energy fraction of PBHs with a mass $M(k)$ over the total DM is given by \cite{Gong:2017qlj, Carr:2016drx}
\begin{equation}\label{fPBH}
f_{\rm{PBH}}(M)\equiv \frac{\Omega_{\rm {PBH}}}{\Omega_{\rm{DM}}}=\frac{\beta(M)}{3.94\times10^{-9}}\left(\frac{\gamma}{0.2}\right)^{1/2}\left(\frac{g_*}{10.75}\right)^{-1/4}\left(\frac{0.12}{\Omega_{\rm{DM}}h^2}\right)
\left(\frac{M}{M_{\odot}}\right)^{-1/2},
\end{equation}
where $\Omega_{\rm {DM}}$ is the current density parameter of DM and its value is constrained by the Planck 2018 results as $\Omega_{\rm {DM}}h^2\simeq0.12$ \cite{akrami2020planck}.

In this work, we focus on three distinct PBH mass scales: ${\cal O}(10)M_\odot$, ${\cal O}(10^{-5})M_\odot$, and ${\cal O}(10^{-13})M_\odot$.


\section{Galileon Inflationary Model}\label{sec:review}

The Galileon inflation is described by the action \cite{Kamada:2010qe,Kobayashi:2010cm}
\begin{equation}\label{action0}
S= \int {\rm d}^{4}x \sqrt{-g}\left[\frac{1}{2} R + {\cal L}_{\phi} \right],
\end{equation}
where $g$ is the determinant of the metric $ g_{{\mu}{\nu}}$, $R$ is the Ricci scalar, and ${\cal L}_{\phi}$ is the scalar field Lagrangian which is given by
\begin{equation}\label{Lagrangian}
{\cal L}_{\phi}\equiv K(\phi, X)-G_3(\phi, X)\Box \phi.
\end{equation}
Here, $K(\phi,X)$ and $G_3(\phi,X)$ are general functions in terms of the Galileon field $\phi$ and the kinetic term $X\equiv -\frac{1}{2}g^{\mu\nu}\partial_\mu{\phi}\partial_\nu{\phi}$. Throughout this paper, we set the reduced Planck mass equal to unity, i.e., $M_{\rm{pl}}=(8\pi G)^{-1/2}=1$. It is worthwhile to mention that, the most general form of the Galileon Lagrangian includes two other terms: $G_4(\phi,X)R+G_{4,X}\times$ [field derivative terms] and $G_5(\phi,X)G^{{\mu}{\nu}}(\bigtriangledown_{\mu}\bigtriangledown_{\nu}\phi) -(G_{5,X}/6)\times$  [field derivative terms], in addition to the term $G_3(\phi, X)\Box \phi$ \cite{Deffayet:2009wt, Deffayet:2009mn}. Here, $G_4(\phi,X)$ and $G_5(\phi,X)$ are functions of $\phi$ and $X$, $G^{{\mu}{\nu}}$ is the Einstein tensor, and $({,X})\equiv{\partial }/{\partial X}$. In the present work, we focus on the case in which $G_3(\phi,X)=G_3(\phi)$ and $G_4=G_5=0$. We also consider $K(\phi, X)=X-V(\phi)$ where $V(\phi)$ is the scalar field potential. With these assumptions and then integration by parts, it is straightforward to find that the action (\ref{action0}) turns to the following form
\begin{equation}\label{action}
S= \int {\rm d}^{4}x \sqrt{-g}\left[\frac{1}{2} R+ \Big(1-2G(\phi)\Big)X-V(\phi)\right],
\end{equation}
where $G(\phi)\equiv dG_3({\phi})/d\phi$. As we see, this model can also be considered as a modified gravitational theory with the noncanonical kinetic term $\omega(\phi)X$ with $\omega(\phi)=1-2G(\phi)$ \cite{de2011chaotic,Rezazadeh:2014fwa,Teimoori:2017wbx}.

Since the Galileon scenario is equivalent to the Horndeski theory, in the following, we follow the approach of \cite{Tsujikawa:2013ila, DeFelice:2011uc} in which we set $K(\phi,X)=X-V(\phi)$, $G_3=G_3(\phi)$, and $G_4=G_5=0$, to review the background evolution and basic formulas governing the theory of cosmological perturbations in our model.

For the flat Friedmann-Robertson-Walker (FRW) metric $g_{{\mu}{\nu}}={\rm diag}\Big(-1, a^{2}(t), a^{2}(t), a^{2}(t)\Big)$, Eqs. (4), (5), (6), and (7) in \cite{DeFelice:2011uc} give the modified background equations as
\begin{align}
\label{eq:FR1}
 & 3 H^{2}-\frac{1}{2}\big(1-2 G(\phi)\big)\dot{\phi}^{2}-V(\phi) =0, \\
  \label{eq:FR2}
 & 2 \dot{H}+3H^2+\frac{1}{2}\big(1-2 G(\phi)\big)\dot{\phi}^{2}-V(\phi) =0,\\
  \label{eq:Field}
 &\big(1-2G(\phi)\big)\big(\ddot{\phi}+3H\dot{\phi}\big)+V_{,\phi}-\dot{\phi}G_{,\phi}=0,
\end{align}
where $H\equiv \dot{a}/a $ is the Hubble parameter. Note that the overdot represents the derivative with respect to the cosmic time $t$, and $({,\phi})\equiv{\partial }/{\partial \phi}$.

Following \cite{ Tsujikawa:2013ila, DeFelice:2011uc}, we introduce the slow-roll parameters as the following forms
\begin{equation}\label{SRparameters}
\varepsilon \equiv -\frac{\dot H}{H^2}, \hspace{.5cm}  \delta_{\phi}\equiv \frac{\ddot{\phi}}{ H\, \dot{\phi}}, \hspace{.5cm}\delta_{X}\equiv \frac{\dot{\phi}^2}{2 H^2}, \hspace{.5cm} \delta_{G}\equiv \frac{G(\phi)\dot{\phi}^2}{2H^2}.
\end{equation}
From the first relation in Eq. (\ref{SRparameters}), it is clear that to have inflation, the condition $\varepsilon<1$ is required. Using Eq. (9) in \cite{DeFelice:2011uc}, the slow-roll parameter $\varepsilon$ can be written as
\begin{equation}\label{epsilon:exact}
\varepsilon =\delta_{X}-2\delta_{G}= \frac{\dot{\phi}^2}{2H^2}\big(1-2G(\phi)\big),
\end{equation}
where in the last equality, we have substituted the third and fourth relation in Eq. (\ref{SRparameters}). Note that, under the slow-roll approximation, all the parameters defined in Eq. (\ref{SRparameters}) should be much smaller than unity.

In the framework of G-inflation, the power spectrum of the scalar perturbation ${\cal P}_{s}$ at the time of sound horizon exit, i.e., $c_{s}k = aH$ in which $k$ is a comoving wavenumber, takes the form \cite{Tsujikawa:2013ila, DeFelice:2011uc}
\begin{equation}\label{Ps0}
{\cal P}_{s}=\frac{H^2}{8 \pi ^{2}Q_{s}c_{s}^3}\Big|_{c_{s}k=aH},
\end{equation}
where the quantities $c_{s}$ and $Q_{s}$ are given by Eqs. (15) and (16) in \cite{Tsujikawa:2013ila}. From Eqs. (15)-(20) in \cite{Tsujikawa:2013ila} with setting $K(\phi,X)=X-V(\phi)$, $G_3=G_3(\phi)$, $G_4=G_5=0$, and also using the first equality in Eq. (\ref{epsilon:exact}), one can easily show that the sound speed $c_{s}$ is equal to the light speed ($c_{s} = 1$) and $Q_s=\delta_{X}-2\delta_{G}=\varepsilon$. Consequently Eq. (\ref{Ps0}) reduces to
\begin{equation}\label{Ps1}
{\cal P}_{s}=\frac{H^2}{8 \pi ^{2}}\frac{1}{\varepsilon}\Big|_{k=aH}.
\end{equation}
Note that, since $c_s^2>0$ and $Q_s>0$, there are no ghosts and Laplacian instabilities for scalar perturbations \cite{DeFelice:2011uc}. The observational value of the amplitude of scalar perturbations at the CMB pivot scale $k_{*}=0.05\,{\rm Mpc}^{{\rm -1}}$ is ${\cal P}_{s}(k_{*})\simeq 2.1 \times 10^{-9}$ \cite{akrami2020planck}.

In the slow-roll approximation, the field equations (\ref{eq:FR1}) and (\ref{eq:Field}) reduce to
\begin{align}
\label{FR1:SR}
& 3 H^2\simeq V(\phi),\\
  \label{Field:SR}
& 3 H\dot{\phi}\big(1-2G(\phi)\big)+V_{,\phi}\simeq0.
\end{align}
Using Eqs. (\ref{FR1:SR}) and (\ref{Field:SR}), the slow-roll parameter $\varepsilon$ in (\ref{epsilon:exact}) can be written as
\begin{equation}\label{epsilon:SR}
\varepsilon \simeq \frac{\varepsilon_{V}}{1-2G(\phi)},
\end{equation}
where
\begin{equation}\label{epsilonv}
\varepsilon_{V}\equiv \frac{1}{2}\left(\frac{V_{,\phi}}{V}\right)^2.
\end{equation}
With the help of Eqs. (\ref{epsilon:SR}), (\ref{epsilonv}), and the first Friedmann Eq. (\ref{FR1:SR}), the power spectrum (\ref{Ps1}) in the slow-roll limit reduces to
\begin{equation}\label{PsSR}
{\cal P}_{s}\simeq \frac{V^3}{12\pi^2 V_{,\phi}^2}\Big(1-2G(\phi)\Big).
\end{equation}
During the slow-roll inflationary phase, the Hubble parameter $H$ and the sound speed $c_s$ change much slower than the scale factor $a$ of the universe \cite{garriga1999perturbations}. Therefore, using the relation $c_s k=a H$, we can write $d\ln k\approx H dt$. Applying this approximation and the definition $n_s-1 \equiv d\ln{\cal P}_{s}/d\ln k$, one can easily find $n_s-1\simeq \dot{{\cal P}_{s}}/(H{\cal P}_{s})$. From this relation and then using Eqs. (\ref{FR1:SR}), (\ref{Field:SR}), (\ref{epsilonv}), and (\ref{PsSR}), we can obtain the scalar spectral index $n_s$ as
\begin{equation}\label{nsSR}
n_s-1\simeq \frac{1}{1-2G(\phi)}\left[2\eta_{V}-6\varepsilon_{V}+\frac{2G_{,\phi}}{1-2G(\phi)}\sqrt{2\varepsilon_{V}}\right],
\end{equation}
where
\begin{equation}\label{etav}
\eta_{V}=\frac{V_{,\phi\phi}}{V}.
\end{equation}
The observational value of the scalar spectral index measured by the Planck team is $n_{s}=0.9627\pm0.0060$ (68\% CL, Planck 2018 TT+lowE) \cite{akrami2020planck}. Applying the approximation $d\ln k\approx H dt$ and Eq. (\ref{Field:SR}), we can find the running of the scalar spectral index as
\begin{equation}\label{alphas}
\frac{d{n_s}}{d\ln k}\simeq -\frac{1}{\big(1-2 G(\phi)\big)}\left(\frac{V_{,\phi}}{V(\phi)}\right) n_{s,\phi}.
\end{equation}
The observational constraint on the running of the scalar spectral index is about $d{n_s}/d \ln k= - {\rm{0}}{\rm{.0078}} \pm 0.0082$ (68\% CL, Planck 2018 TT+lowE) \cite{akrami2020planck}.

The tensor power spectrum in the framework of G-inflation is given by \cite{Tsujikawa:2013ila, DeFelice:2011uc}
\begin{equation}\label{Pt0}
{\cal P}_{t}=\frac{H^2}{2\pi ^{2}Q_{t}c_{t}^{3}}\Big|_{c_{t}k=aH},
\end{equation}
where the quantities $Q_t$ and $c_{t}$ can be found with the help of Eqs. (17)-(20), (31), and (32) in \cite{Tsujikawa:2013ila}, by setting $K(\phi,X)=X-V(\phi)$, $G_3=G_3(\phi)$, and $G_4=G_5=0$ as $Q_t=\frac{1}{4}$ and $c_{t}^2=1$. In this way, the tensor power spectrum is derived as
\begin{equation}\label{Pt1}
{\cal P}_{t}=\frac{H^2}{2\pi ^{2}}\Big|_{k=aH},
\end{equation}
which is the same as one in the standard canonical inflationary model. As we see, the requirements to avoid the ghosts ($Q_t>0$) and Laplacian ($c_t^2>0$) instabilities are fulfilled for the tensor perturbations too \cite{DeFelice:2011uc}.

For our Galileon model described by Eq. (\ref{action}), since $c_s^2=1$, with the help of Eqs. (24) and (34) in \cite{Tsujikawa:2013ila}, Eq. (\ref{epsilonv}) and also the first equality in Eq. (\ref{epsilon:SR}), we get the tensor-to-scalar ratio in the slow-roll regime as
\begin{equation}\label{r}
r\simeq 16 \varepsilon = \frac{16\varepsilon_V}{1-2G(\phi)}.
\end{equation}
The tensor-to-scalar ratio is constrained by the Planck 2018 data as  $r< 0.0654$ (68\% CL, Planck 2018 TT+lowE) \cite{akrami2020planck}.


\section{Primordial Black Holes Formation from $\alpha$-attractor G-inflation} \label{sec:PBHGalileon}

In this section, we discuss about the possibility of PBHs production in the framework of G-inflation described by Eq. (\ref{action}). A careful look to Eq. (\ref{PsSR}) shows that a peak is created in the power spectrum of the curvature perturbations, when the term $\big(1-2G(\phi)\big)$ has a sufficiently large peak about a field value. On the other hand, the term $\big(1-2G(\phi)\big)$ becomes maximum when the function $-G(\phi)$ maximizes. To achieve an appropriate peak on a particular scale in the scalar power spectrum which can lead to form PBHs, we take the functional form of $G(\phi)$ as \cite{Kallosh:2013yoa, Kallosh:2014rga, Kallosh:2014laa, Kallosh:2015lwa, Linde:2015uga, Galante:2014ifa, Fu:2019ttf}
\begin{equation}\label{G}
G(\phi)=G_I(\phi)\Big(1+G_{II}(\phi)\Big),
\end{equation}
where
\begin{equation}\label{GI}
G_I(\phi)=\frac{1}{2}\left(1-\frac{2\alpha}{(1-\frac{\phi^2}{3})^2}\right),
\end{equation}
and
\begin{equation}\label{GII}
G_{II}(\phi)=\frac{\omega}{\sqrt{\left(\frac{\phi-\phi_c}{\sigma}\right)^2+1}}.
\end{equation}
As we see, the Galileon term contains two functions $G_I(\phi)$ and $G_{II}(\phi)$. The quantity $G_{II}(\phi)$ is a function which has a peak at the critical field value $\phi = \phi_c$, and the height and width of the peak are determined by the parameters $\omega$ and $\sigma$, respectively. The function $G_{II}(\phi)$ almost vanishes for the field values away from $\phi_c$, so that $G(\phi)\approx G_I(\phi)$. In such a way, the kinetic term in the action (\ref{action}) nearly reduces to $\Big(2\alpha X/(1-\phi^{2}/3)^2\Big)$ which is the same as one in $\alpha$-attractors \cite{Kallosh:2013yoa,Galante:2014ifa}. Consequently, our G-inflation model described by Eq. (\ref{action}) is transformed to the $\alpha$-attractor models at significant period of time domain of interest during inflation. This means that the inflationary observables $n_s$ and $r$, more likely are in agreement with the observational data at around the sound horizon exit. Because, the $\alpha$-attractor models for the large values of the $e$-fold number $N$ and small $\alpha$, have the same prediction for $n_s$ and $r$ in the Einstein frame as $n_s=1-2/N$ and $r=12\alpha/N^2$ which for $\alpha=1$ these quantities are transformed to the corresponding results in the Starobinsky model \cite{Starobinsky:1980te} and also in the Higgs inflationary scenario \cite{Bezrukov:2007ep}. Here, $\phi_c$ and $\sigma$ have dimensions of mass and $\omega$ is a dimensionless parameter. Fine-tuning of these parameters can produce a sufficiently large peak in the curvature power spectrum to form PBHs within a particular mass window.

The simplest choice for the potential of the scalar field in the $\alpha$-attractor models is a power-law function as
\begin{equation}\label{potential}
V(\phi)=V_{0}\phi^{2n},
\end{equation}
where $V_{0}$ and $n$ are constants. By redefining $\phi/\sqrt{3}=1-e^{-\sqrt{\frac{2}{3\alpha}}\varphi}$, the kinetic term $[2\alpha X/(1-\phi^{2}/3)^2]$ reduces to the canonical form $-g^{\mu\nu}\partial_{\mu}\varphi\partial_{\mu}\varphi/2$. This class of the $\alpha$-attractors is called E-models \cite{Kallosh:2013yoa,Galante:2014ifa,Eshaghi:2016kne}. The potential (\ref{potential}) for the scalar field $\varphi$ takes the form $V(\varphi)=\tilde{V_0}(1-e^{-\sqrt{\frac{2}{3\alpha}}\varphi})^{2n}$, where $\tilde{V_0}$ is a constant, and for $n=1$ and $\alpha=1$ it covers the potential of the Starobinsky model in the Einstein frame \cite{Starobinsky:1980te}. In this work, we choose $n=1$ and $\alpha=1$, and then investigate the possibility of PBHs formation for the three parameter sets that are listed in Table \ref{Table1}.

\begin{table}[ht!]
  \centering
  \caption{The chosen parameter sets for $G_{II}(\phi)$ that can successfully produce PBHs. The value of $V_0$ is fixed by imposing the CMB normalization at the pivot scale $k_{*}=0.05\,{\rm Mpc}^{{\rm -1}}$ corresponding to $N_*=60$.}
\scalebox{1}[1] {
    \begin{tabular}{ccccc}
    \hline
    \hline
    $\qquad \# \qquad$ & $\qquad \phi_{c} \qquad$ & $\qquad \qquad \omega \qquad \qquad$ & $\qquad \qquad \sigma \qquad \qquad$ & $\qquad V_{0}\qquad$\tabularnewline
    \hline
 Case 1 & $1.657096$ & $7.86618\times10^{7}$ & $3.3918\times10^{-11}$ & $8.1108\times10^{-11}$\tabularnewline
    \hline
 Case 2 & $1.678000$ & $6.42000\times10^{7}$ & $2.1000\times10^{-11}$ & $8.2000\times10^{-11}$\tabularnewline
    \hline
  Case 3 & $1.689490$ & $6.36100\times10^{7}$ & $1.3000\times10^{-11}$ & $7.6000\times10^{-11}$\tabularnewline
    \hline
    \end{tabular}
    }
  \label{Table1}
\end{table}

To find the value of $V_0$ which represents the energy scale of inflation, we fix the power spectrum in Eq. (\ref{PsSR}) at the pivot scale $k_{*}=0.05\,{\rm Mpc}^{{\rm -1}}$ as ${\cal P}_{s}(k_{*})\simeq 2.1 \times 10^{-9}$ \cite{akrami2020planck}. We calculate the inflationary observables $n_s$, $r$, and $dn_s/d\ln k$ at the $k_{*}=0.05\,{\rm Mpc}^{{\rm -1}}$ and also quantities relevant for producing PBHs of these three parameter sets. The results are summarized in Table \ref{Table2}.

\begin{table*}[ht!]
  \centering
  \caption{Results of the scalar power spectrum at peak scale $({\cal P}_s^{\rm{peak}})$, the mass of the corresponding PBHs $(M_{\rm{PBH}}^{\rm{peak}})$, PBHs fractional abundance $(f_{\rm{PBH}}^{\rm{peak}})$, and also the inflationary observables $n_s$, $r$, and $dn_s/d\ln k$ at the CMB scale, for the three cases of Table \ref{Table1}.
  }
\scalebox{1}[1] {\begin{tabular}{c c c c c c c}
    \hline
    \hline
     \#  & \,\,\,\,\,\,\,\,\,$n_s$ &\,\,\,\,\,\,\,\,\,  $r$  &\,\,\,\,\,\,\,\,\,  $dn_s/d\ln k$   &\,\,\,\,\,\,\,\,\,${\cal P}_s^{\rm{peak}}$ &\,\,\,\,\,\,\,\,\, $M_{\rm{PBH}}^{\rm{peak}}/M_\odot$ &\,\,\,\,\,\,\,\,\, $f_{\rm{PBH}}^{\rm{peak}}$\\
    \hline
    Case 1 &\,\,\,\,\,\,\,\,\, $0.9510$ &\,\,\,\,\,\,\,\,\, $0.0076$& \,\,\,\,\,\,\,\,\, $-0.0011$  & \,\,\,\,\,\,\,\,\,$0.0319$ &\,\,\,\,\,\,\,\,\,$8.06\times10^{-13}$ &\,\,\,\,\,\,\,\,\,$0.9750$\\
    Case 2 &\,\,\,\,\,\,\,\,\, $0.9521$ &\,\,\,\,\,\,\,\,\, $0.0077$&\,\,\,\,\,\,\,\,\,  $-0.0008$  &\,\,\,\,\,\,\,\,\,$0.0379$  &\,\,\,\,\,\,\,\,\,$1.78\times10^{-5}$ &\,\,\,\,\,\,\,\,\, $0.0178$\\
   Case 3 &\,\,\,\,\,\,\,\,\, $0.9586$ &\,\,\,\,\,\,\,\,\, $0.0074$& \,\,\,\,\,\,\,\,\,  $0.0007$   & \,\,\,\,\,\,\,\,\,$0.0462$ &\,\,\,\,\,\,\,\,\,$12.99$ &\,\,\,\,\,\,\,\,\,$0.0020$\\
    \hline
    \end{tabular}}
  \label{Table2}
\end{table*}

\begin{figure*}
\begin{minipage}[b]{1\textwidth}
\subfigure{\includegraphics[width=.48\textwidth]%
{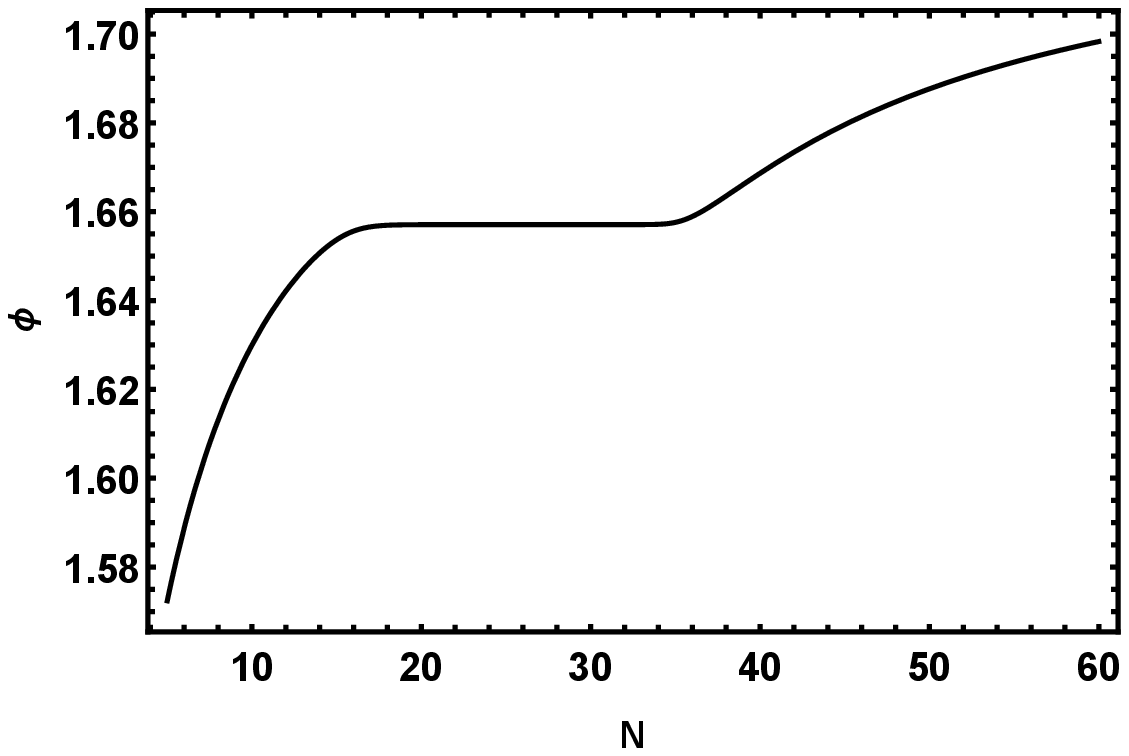}} \hspace{.1cm}
\subfigure{ \includegraphics[width=.48\textwidth]%
{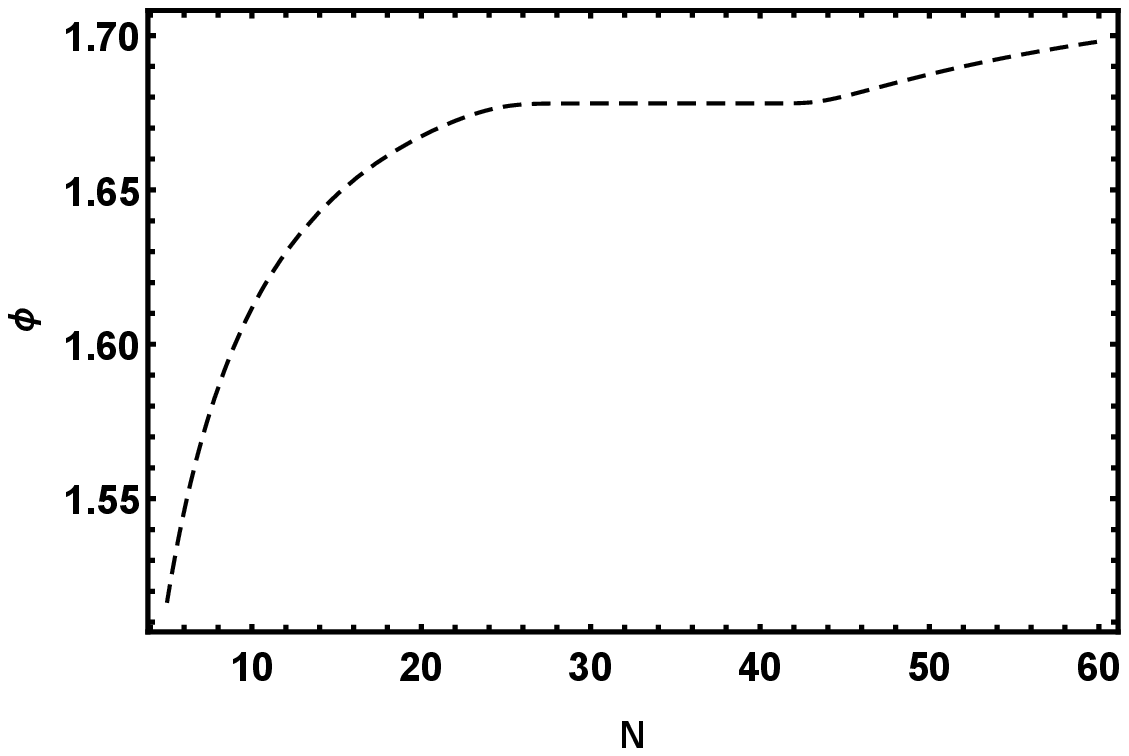}}\hspace{.1cm}
\subfigure{ \includegraphics[width=.48\textwidth]%
{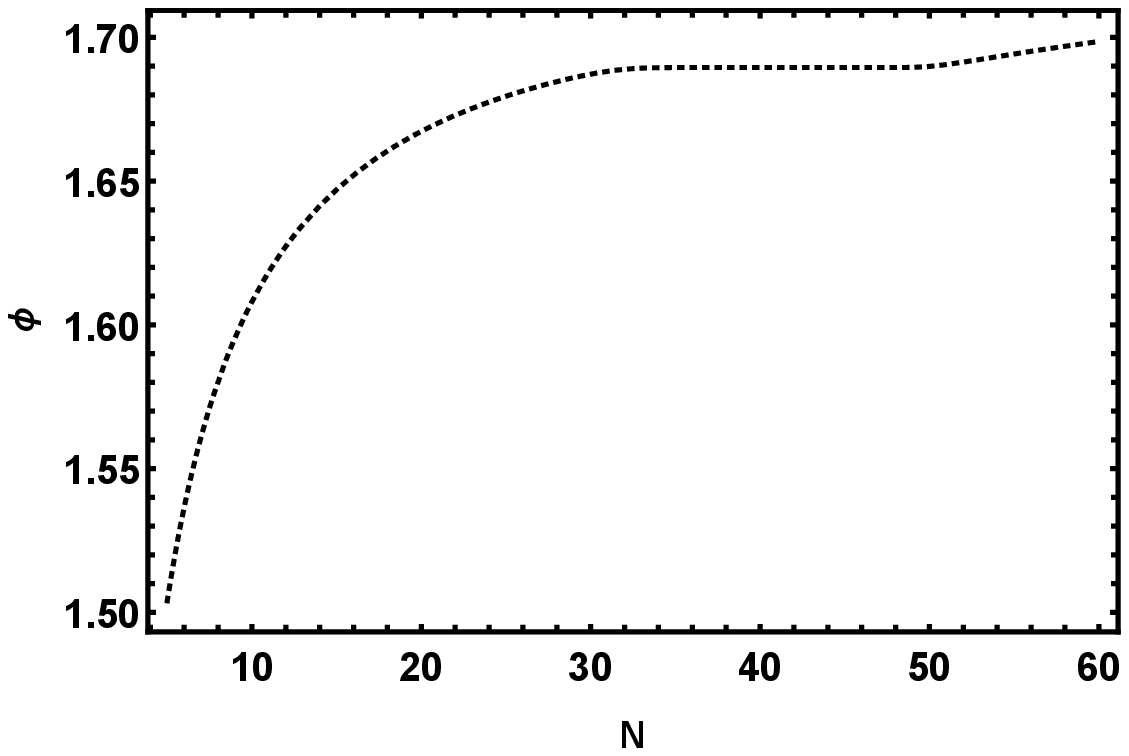}}
\end{minipage}
\caption{Evolution of the scalar field $\phi$ as a function of the $e$-fold number $N$ for Case 1 (solid line), Case 2 (dashed line), and Case 3 (dotted line). The initial conditions are found by using Eqs. (\ref{FR1:SR}) and (\ref{Field:SR}) at $N_{*}=60$.}
\label{fig:phiN}
\end{figure*}

In Fig. \ref{fig:phiN}, we plot the evolution of the scalar field $\phi$ as a function of the $e$-fold number $N$ where $dN=-Hdt$, for parameter set 1 (solid line),  parameter set 2 (dashed line), and parameter set 3 (dotted line), by solving the background equations (\ref{eq:FR1})-(\ref{eq:Field}) numerically. The initial conditions are set by using Eqs. (\ref{FR1:SR}) and (\ref{Field:SR}) at $N_{*}=60$. As we see, in this figure there are plateau-like regions at $\phi=\phi_c$ corresponding to $17 \lesssim N\lesssim 35$ for parameter set 1, $24 \lesssim N\lesssim 43$ for  parameter set 2, and $31\lesssim N\lesssim 50$ for parameter set 3. During this region, inflaton experiences an ultra slow-roll phase in which its velocity rapidly decreases and the curvature power spectrum can be enhanced by several orders of magnitude which can lead to PBHs formation. Besides, we see in the figure that displacements of the inflaton field from the epoch of horizon crossing to the end of inflation ($\Delta\phi\equiv\left|\phi_{*}-\phi_{\mathrm{end}}\right|$) in our setup are obtained as $0.340M_{\rm{pl}}$, $0.868M_{\rm{pl}}$, and $1.124M_{\rm{pl}}$ for Case 1, Case 2, and Case 3, respectively. Therefore, since the field excursion in our model takes sub-Planckian values for Case 1 and Case 2, the distance swampland conjecture \cite{Obied:2018sgi, Garg:2018reu,Ooguri:2018wrx} is preserved for these two parameter sets while for Case 3, the result is super-Planckian which violates the distance swampland conjecture. In the original $\alpha$-attractor inflation model \cite{Kallosh:2013yoa, Kallosh:2014rga, Kallosh:2014laa, Kallosh:2015lwa, Linde:2015uga, Galante:2014ifa, Carrasco:2015rva, Carrasco:2015pla, Roest:2015qya, scalisi2015cosmological, Eshaghi:2016kne} which can be recovered by setting $G_{II}(\phi)=0$ in our scenario, and in its noncanonical frame, the field excursion is obtained as $0.920 M_{\rm{pl}}$ which is sub-Planckian and preserves the distance swampland conjecture \cite{Obied:2018sgi, Ooguri:2018wrx}, but in the Einstein frame, we obtain the field excursion as $4.513M_{\rm{pl}}$ which is super-Planckian and hence violates the required condition.

\begin{figure*}[t]
\begin{center}
\scalebox{1}[1]{\includegraphics{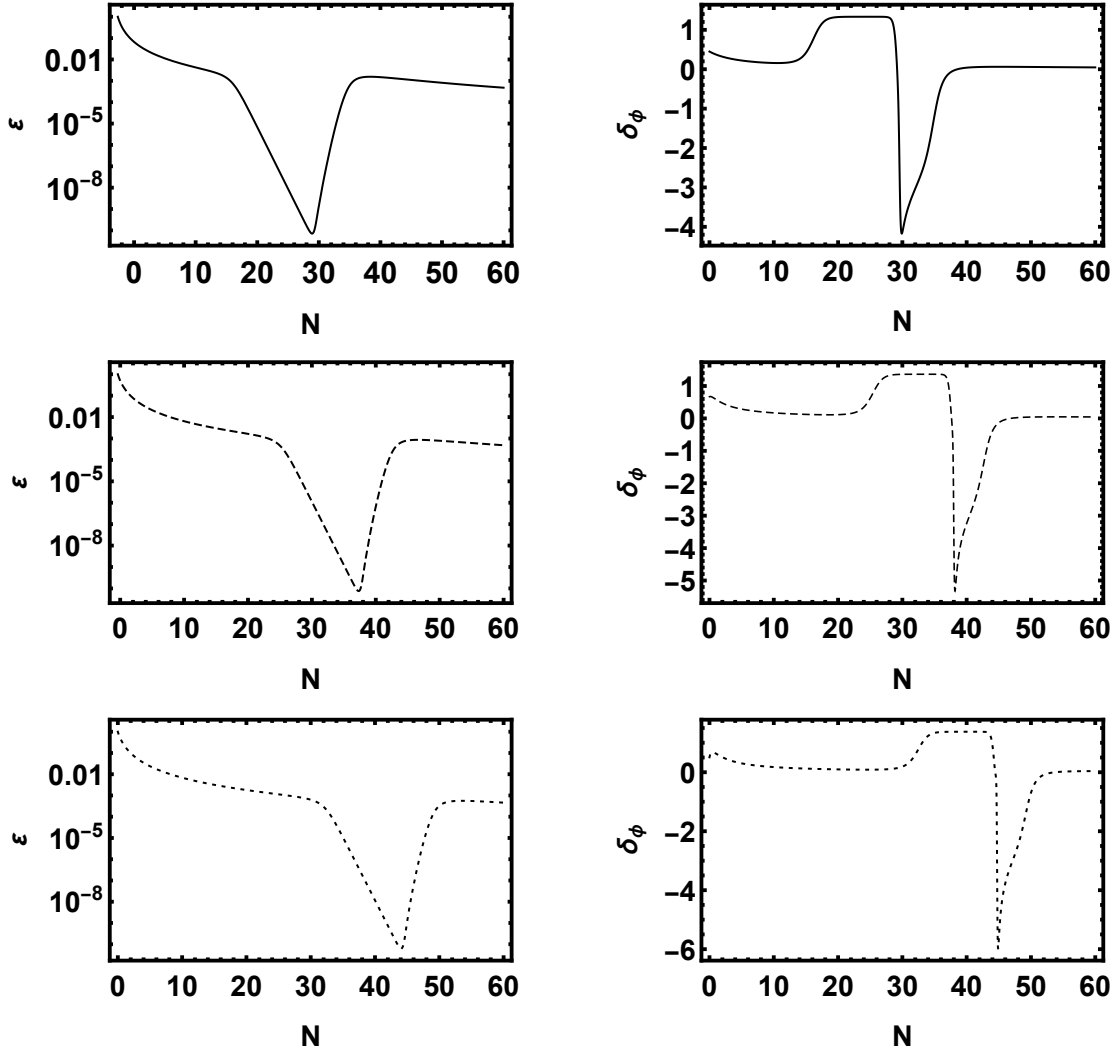}}
\caption{Evolution of the first slow-roll parameter $\varepsilon$ (left) and the second slow-roll parameter $\delta_{\phi}$ (right) versus the $e$-fold number $N$ for Case 1 (solid line), Case 2 (dashed line), and Case 3 (dotted line).}
\label{fig:SRparameters}
\end{center}
\end{figure*}

In Fig. \ref{fig:SRparameters}, the evolution of the slow-roll parameters $\varepsilon$ (left panel) and $\delta_\phi$ (right panel) are plotted as functions of $N$ for the three  parameter sets of Table \ref{Table1}. The left panel of Fig. \ref{fig:SRparameters} shows that $\varepsilon$ remains below unity during the whole inflationary period $(\varepsilon<1)$, but in the right panel, we see that $\delta_{\phi}$ breaks the slow-roll approximation during the ultra slow-roll stage. It is worth mentioning that in the slow-roll approximation both parameters $\varepsilon$ and $\delta_\phi$ should be less than unity. Of course, from the left panel of Fig. \ref{fig:SRparameters}, we see that $\varepsilon$ can not reach unity at the end of inflation, i.e., $N=0$ and inflation continues 2.640 $e$-folds for parameter set 1, 0.372 $e$-folds for parameter set 2, and 0.004 $e$-folds for parameter set 3, due to strong slow down of inflaton during the ultra slow-roll stage.

From Fig. \ref{fig:SRparameters}, it is clear that at the time of sound horizon exit corresponding to $N_{*}=60$, the slow-roll approximation remains valid. Therefore, with the help of Eqs. (\ref{nsSR}), (\ref{alphas}), (\ref{r}) and also using Eqs. (\ref{epsilonv}), (\ref{etav}), and (\ref{G})-(\ref{potential}), we can find the values of scalar spectral index $n_s$, the tensor-to-scalar ratio $r$, and the running of the scalar spectral index $dn_s/d\ln k$ for the three parameter sets of Table \ref{Table1}. The numerical results are presented in Table \ref{Table2} and imply that the values of $n_s$ for  parameter sets 1 and 2 satisfy the $95\%$ CL constraints of Planck 2018 TT+lowE data \cite{akrami2020planck}, and the values of $dn_s/d\ln k$ and $r$ are in agreement with the $68\%$ CL constraints of these data. For parameter set 3, the values of $n_s$ and $r$ are consistent with the $68\%$ CL constraints of Planck 2018 TT+lowE data \cite{akrami2020planck}, while the value of $dn_s/d\ln k$ is in agreement with the $95\%$ CL constraints of these data. It is worth to mention that from the obtained numerical results in Table \ref{Table2}, it is realized that our model predicts low values for the tensor-to-scalar ratio $r$. This important remark makes the model very interesting.

With these considerations, using Eq. (\ref{PsSR}) (or Eq. (\ref{Ps1})) to compute the scalar power spectrum is not very accurate and gives wrong results. To obtain the exact power spectrum, we should solve the Mukhanov-Sasaki equation numerically for all the Fourier modes of interest. Since, the sound speed of the scalar perturbation in our model is equal to the light speed, i.e., $c_s=1$, the Mukhanov-Sasaki equation takes the form
\begin{equation}\label{MS}
 \upsilon^{\prime\prime}_{k}+\left(k^2-\frac{z^{\prime\prime}}{z}\right)\upsilon_k=0,
\end{equation}
where the prime represents derivative with respect to the conformal time $\eta=\int {a^{-1}dt}$ and
\begin{equation}\label{z}
 \upsilon\equiv z {\cal R}, \hspace{1cm} z=a\sqrt{2Q_s}.
\end{equation}
The Mukhanov-Sasaki equation (\ref{MS}) describes the evolution of the Fourier modes $\upsilon_k$. Each mode $\upsilon_k$ evolves during inflation, until it exits the Hubble horizon and approaches a constant value. By solving the Mukhanov-Sasaki equation numerically, we find the evolution of real and imaginary parts of $\upsilon_k$, then we estimate the scalar power-spectrum of each mode $\upsilon_k$ using the following relation
\begin{equation}\label{PsBunch}
{\cal P}_{s}=\frac{k^{3}}{2\pi^{2}}\left|\frac{\upsilon_{k}^{2}}{z^{2}}\right|_{k\ll aH}.
\end{equation}
The initial conditions for each mode $\upsilon_k$ are determined by assuming that when each mode is sub-horizon $(k\gg aH)$, $\upsilon_k$ is in the Bunch-Davies vacuum as \cite{DeFelice:2013ar}
\begin{equation}\label{Bunch-Davies}
\upsilon_k=\frac{e^{-ik\tau}}{\sqrt{2 k}}.
\end{equation}

\begin{figure}[t]
\begin{center}
\scalebox{0.9}[0.9]{\includegraphics{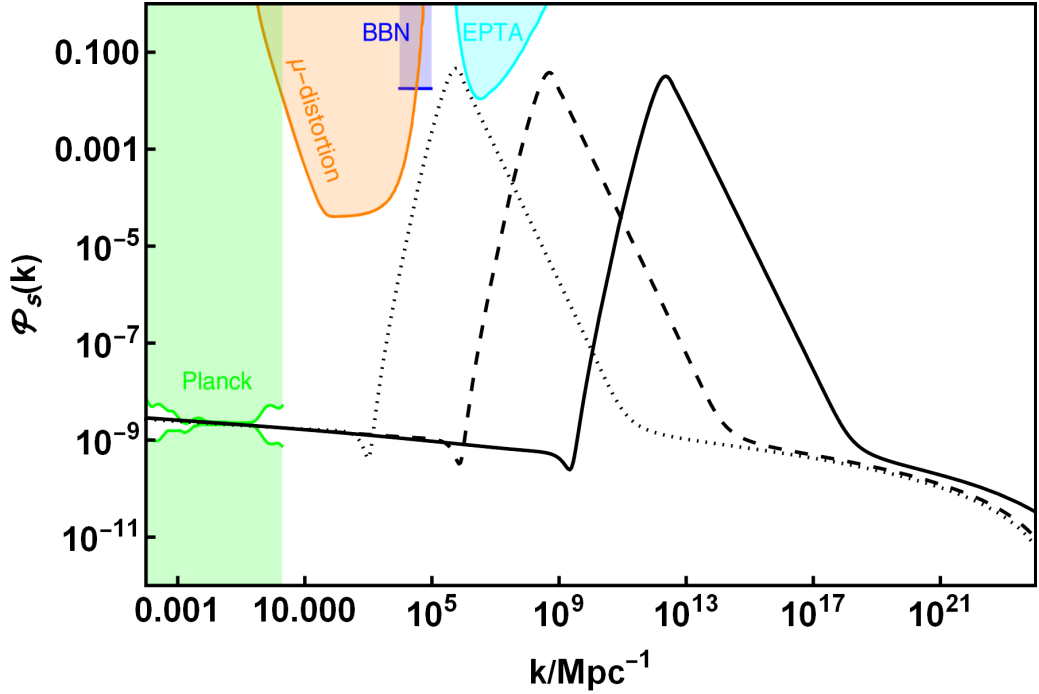}}
\caption{The curvature power spectra calculated by solving the Mukhanov-Sasaki equation numerically as a function of the comoving wavenumber $k$ for Case 1 (solid), Case 2 (dashed) and Case 3 (dotted). The light green shaded region shows the area excluded by the CMB observations \cite{akrami2020planck}. The orange, blue, and cyan shaded regions represent the excluded regions for the power spectrum by the $\mu$-distortion of CMB \cite{Fixsen:1996nj, Chluba:2012we}, the effect on the ratio between neutron and proton during the big bang nucleosynthesis (BBN) \cite{Jeong:2014gna, Inomata:2016uip}, and the current PTA observations \cite{Inomata:2018epa}, respectively.}
\label{psk}
\end{center}
\end{figure}

In Fig. \ref{psk}, we plot the power spectra of the curvature perturbations computed by solving the Mukhanov-Sasaki equation (\ref{MS}) numerically as a function of the comoving wavenumber $k$, for the three parameter sets 1, 2 and 3. From this figure, it is obvious that on the large scales which the scalar field experiences a slow-roll phase, the three curves are in good agreement with the CMB constraints \cite{akrami2020planck}. We also see, when the inflaton enters the ultra slow-roll phase, the power spectra get amplified to the order of ${\cal O}(10^{-2})$ which is large enough to have a significant PBH abundance.

\begin{figure}[t]
\begin{center}
\scalebox{0.9}[0.9]{\includegraphics{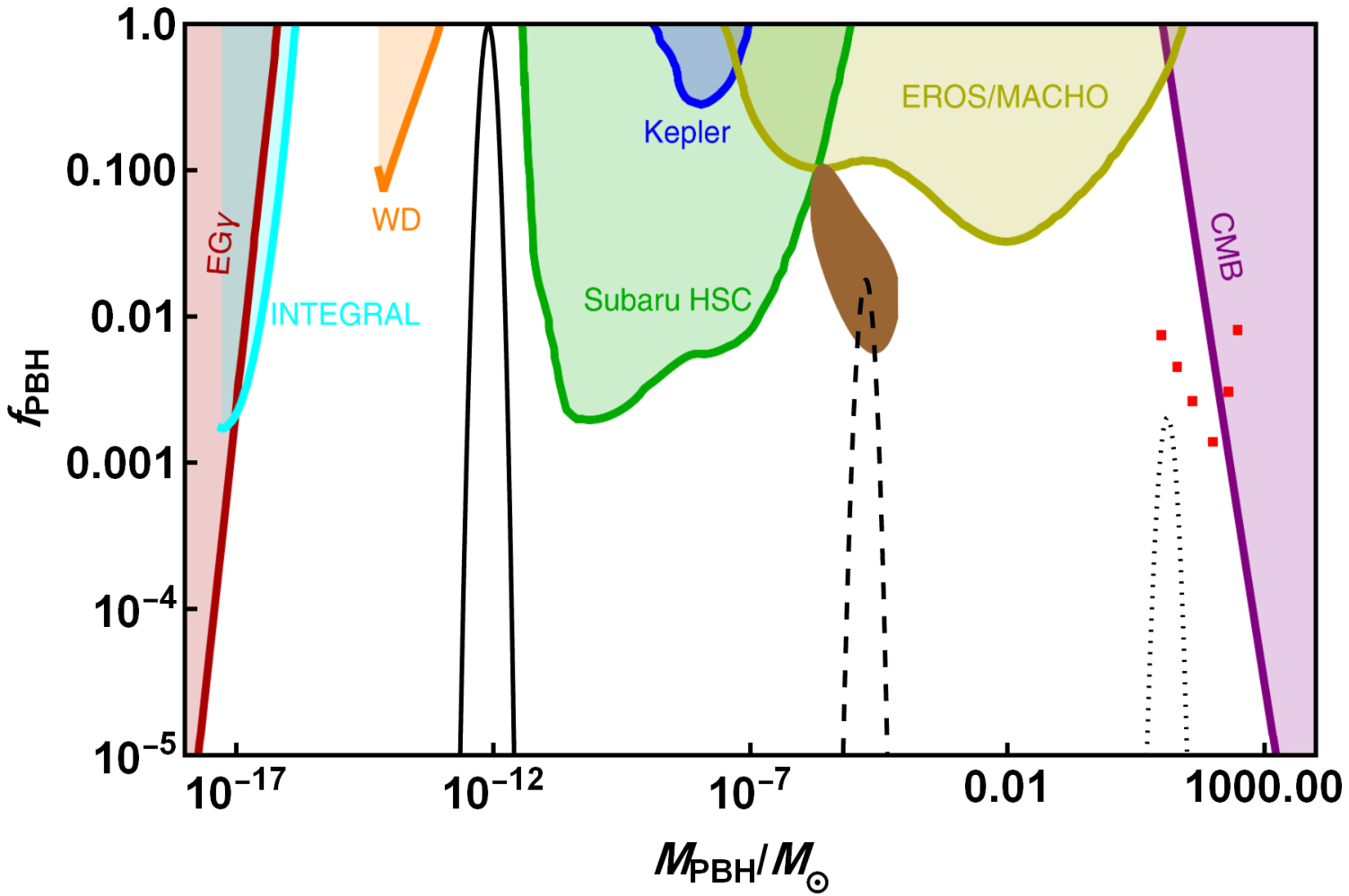}}
\caption{The fractional abundance of PBHs as a function of PBH mass for Case 1 (solid line), Case 2 (dashed line), and Case 3 (dotted line). The red points represents the potential upper bounds on the PBH abundance from requiring that the merger rate of PBHs does not exceed the upper limit on the LIGO merger rate \cite{Ali-Haimoud:2017rtz}. The brown-shaded region shows the allowed PBH abundance from the ultrashort-timescale microlensing events in the OGLE data \cite{mroz2017no, Niikura:2017zjd}. The other shaded regions indicate the current observational constraints on the abundance of PBHs: extragalactic gamma-rays from PBH evaporation (EG$\gamma$) \cite{Carr:2009jm}, galactic center 511 keV $\gamma$-ray line (INTEGRAL) \cite{Laha:2019ssq}, white dwarfs explosion (WD) \cite{Graham:2015apa}, microlensing events with Subaru HSC (Subaru HSC) \cite{Niikura:2017zjd}, with the Kepler satellite (Kepler) \cite{Griest:2013esa}, with EROS/MACHO (EROS/MACHO) \cite{Tisserand:2006zx}, and accretion constraints from CMB (CMB) \cite{Ali-Haimoud:2016mbv}.
}
\label{fPBH-M}
\end{center}
\end{figure}

With the help of actual power spectrum obtained by solving the Mukhanov-Sasaki equation numerically and using Eqs. (\ref{PBHmass})-(\ref{fPBH}), we can compute PBHs abundance for parameter sets 1, 2, and 3. The results are shown in Fig. \ref{fPBH-M} and Table \ref{Table2}. For parameter set 1, our model predicts PBHs with mass $M\simeq 8.06\times10^{-13} M_\odot$ and PBH abundance $f_{\rm{PBH}}\simeq 0.9750$, which means that the formed PBHs in this class constitute around $0.98\%$ of DM, and therefore, it can be an attractive candidate for DM.

For parameter set 2, the model predicts PBHs with the mass $1.78\times10^{-5}M_\odot$ and $f_{\rm {PBH}}\simeq 0.0178$. From Fig. \ref{fPBH-M}, we see that the peak of $f_{\rm PBH}$ places on the inferred region of the PBH abundance by the ultrashort-timescale microlensing events in OGLE data. Thus, one can take these PBHs as a source of these microlensing events.

For parameter set 3, the model generates PBHs with mass $M\simeq 12.99 M_\odot$ and $f_{\rm {PBH}}\simeq 0.0020$. Fig. \ref{fPBH-M} shows that the produced PBHs in this class satisfy the constraint from the upper limit on the LIGO merger rate.


\section{Secondary gravitational waves}
\label{sec:sgws}

The enhancement of the power spectrum of the curvature perturbations at some scales during inflation may also induce the production of secondary GWs. Recently several earth-based or space-based observatories have been designed that may detect the signal of this type of GWs, if their present fractional energy density lies within the sensitivity regions of the detectors. Therefore, due to their observational capabilities, the secondary GWs have attracted substantial attention among the inflationary contexts recently. In this section, we compute the secondary GWs in the setup of our $\alpha$-attractor G-inflation, and compare its predictions with the sensitivity regions of some designed GWs detectors.

The fractional energy density of the induced GWs in the radiation dominated (RD) era is given by \cite{Kohri:2018awv, Lu:2019sti}
\begin{equation}
 \label{OmegaGW}
 \Omega_{\rm GW}(k,\eta)=\frac{1}{6}\left(\frac{k}{aH}\right)^{2}\int_{0}^{\infty}dv\int_{|1-v|}^{|1+v|}du\left(\frac{4v^{2}-\left(1-u^{2}+v^{2}\right)^{2}}{4uv}\right)^{2}\overline{I_{RD}^{2}(u,v,x)}\mathcal{P}_{s}(ku)\mathcal{P}_{s}(kv),
\end{equation}
where $\eta$ denotes the conformal time, and the time average of the source terms is implied by
\begin{align}
 \overline{I_{\rm RD}^{2}(u,v,x\to\infty)}= & \frac{1}{2x^{2}}\Bigg[\left(\frac{3\pi\left(u^{2}+v^{2}-3\right)^{2}\Theta\left(u+v-\sqrt{3}\right)}{4u^{3}v^{3}}+\frac{T_{c}(u,v,1)}{9}\right)^{2}
 \nonumber\\
 & +\left(\frac{\tilde{T}_{s}(u,v,1)}{9}\right)^{2}\Bigg].
 \label{IRD2b}
\end{align}
In the above equation, we have defined the following functions
\begin{align}
T_{c}= & -\frac{27}{8u^{3}v^{3}x^{4}}\Bigg\{-48uvx^{2}\cos\left(\frac{ux}{\sqrt{3}}\right)\cos\left(\frac{vx}{\sqrt{3}}\right)\left(3\sin(x)+x\cos(x)\right)+
\nonumber\\
& 48\sqrt{3}x^{2}\cos(x)\left(v\sin\left(\frac{ux}{\sqrt{3}}\right)\cos\left(\frac{vx}{\sqrt{3}}\right)+u\cos\left(\frac{ux}{\sqrt{3}}\right)\sin\left(\frac{vx}{\sqrt{3}}\right)\right)+
\nonumber\\
& 8\sqrt{3}x\sin(x)\Bigg[v\left(18-x^{2}\left(u^{2}-v^{2}+3\right)\right)\sin\left(\frac{ux}{\sqrt{3}}\right)\cos\left(\frac{vx}{\sqrt{3}}\right)+
\nonumber\\
& u\left(18-x^{2}\left(-u^{2}+v^{2}+3\right)\right)\cos\left(\frac{ux}{\sqrt{3}}\right)\sin\left(\frac{vx}{\sqrt{3}}\right)\Bigg]+
\nonumber\\
& 24x\cos(x)\left(x^{2}\left(-u^{2}-v^{2}+3\right)-6\right)\sin\left(\frac{ux}{\sqrt{3}}\right)\sin\left(\frac{vx}{\sqrt{3}}\right)+
\nonumber\\
& 24\sin(x)\left(x^{2}\left(u^{2}+v^{2}+3\right)-18\right)\sin\left(\frac{ux}{\sqrt{3}}\right)\sin\left(\frac{vx}{\sqrt{3}}\right)\Bigg\}
\nonumber\\
& -\frac{\left(27\left(u^{2}+v^{2}-3\right)^{2}\right)}{4u^{3}v^{3}}\Bigg\{\text{Si}\left[\left(\frac{u-v}{\sqrt{3}}+1\right)x\right]-\text{Si}\left[\left(\frac{u+v}{\sqrt{3}}+1\right)x\right]
\nonumber\\
& +\text{Si}\left[\left(1-\frac{u-v}{\sqrt{3}}\right)x\right]-\text{Si}\left[\left(1-\frac{u+v}{\sqrt{3}}\right)x\right]\Bigg\},
\label{Tc}
\end{align}

\begin{align}
T_{s}= & \frac{27}{8u^{3}v^{3}x^{4}}\Bigg\{48uvx^{2}\cos\left(\frac{ux}{\sqrt{3}}\right)\cos\left(\frac{vx}{\sqrt{3}}\right)\left(x\sin(x)-3\cos(x)\right)-
\nonumber\\
& 48\sqrt{3}x^{2}\sin(x)\left(v\sin\left(\frac{ux}{\sqrt{3}}\right)\cos\left(\frac{vx}{\sqrt{3}}\right)+u\cos\left(\frac{ux}{\sqrt{3}}\right)\sin\left(\frac{vx}{\sqrt{3}}\right)\right)+
\nonumber\\
& 8\sqrt{3}x\cos(x)\Bigg[v\left(18-x^{2}\left(u^{2}-v^{2}+3\right)\right)\sin\left(\frac{ux}{\sqrt{3}}\right)\cos\left(\frac{vx}{\sqrt{3}}\right)+
\nonumber\\
& u\left(18-x^{2}\left(-u^{2}+v^{2}+3\right)\right)\cos\left(\frac{ux}{\sqrt{3}}\right)\sin\left(\frac{vx}{\sqrt{3}}\right)\Bigg]+
\nonumber\\
& 24x\sin(x)\left(6-x^{2}\left(-u^{2}-v^{2}+3\right)\right)\sin\left(\frac{ux}{\sqrt{3}}\right)\sin\left(\frac{vx}{\sqrt{3}}\right)+
\nonumber\\
& 24\cos(x)\left(x^{2}\left(u^{2}+v^{2}+3\right)-18\right)\sin\left(\frac{ux}{\sqrt{3}}\right)\sin\left(\frac{vx}{\sqrt{3}}\right)\Bigg\}-\frac{27\left(u^{2}+v^{2}-3\right)}{u^{2}v^{2}}+
\nonumber\\
& \frac{\left(27\left(u^{2}+v^{2}-3\right)^{2}\right)}{4u^{3}v^{3}}\Bigg\{-\text{Ci}\left[\left|1-\frac{u+v}{\sqrt{3}}\right|x\right]+\ln\left|\frac{3-(u+v)^{2}}{3-(u-v)^{2}}\right|+
\nonumber\\
& \text{Ci}\left[\left(\frac{u-v}{\sqrt{3}}+1\right)x\right]-\text{Ci}\left[\left(\frac{u+v}{\sqrt{3}}+1\right)x\right]+\text{Ci}\left[\left(1-\frac{u-v}{\sqrt{3}}\right)x\right]\Bigg\}.
\label{Ts}
\end{align}
Additionally, the sine-integral $\text{Si}(x)$ and cosine-integral $\text{Ci}(x)$ functions are defined respectively as follows
\begin{equation}
 \label{SiCi}
 \text{Si}(x)=\int_{0}^{x}\frac{\sin(y)}{y}dy,\qquad\text{Ci}(x)=-\int_{x}^{\infty}\frac{\cos(y)}{y}dy.
\end{equation}
The function $\tilde{T}_{s}(u,v,1)$ which also appears in Eq. \eqref{IRD2b} is defined as
\begin{equation}
 \label{Tst}
 \tilde{T}_{s}(u,v,1)=T_{s}(u,v,1)+\frac{27\left(u^{2}+v^{2}-3\right)}{u^{2}v^{2}}-\frac{27\left(u^{2}+v^{2}-3\right)^{2}}{4u^{3}v^{3}}\ln\left|\frac{3-(u+v)^{2}}{3-(u-v)^{2}}\right|.
\end{equation}
The present-day energy density fraction of the included GWs has the following relation with the corresponding values well after their horizon re-entry in the radiation domination epoch,
\begin{equation}
 \label{OmegaGW0}
 \Omega_{\rm GW}\left(k,\eta_{0}\right)=\Omega_{\rm GW}(k,\eta)\frac{\Omega_{\rm r_0}}{\Omega_{\rm r}(\eta)},
\end{equation}
where $\Omega_{\rm r}$ indicates the fractional energy density of radiation, and the subscript $0$ refers to the present epoch. Here, we adopt the present-day radiation density parameter to be $\Omega_{\rm r_0}h^{2}\simeq4.2\times10^{-5}$ \cite{Cai:2019bmk, Fu:2019vqc, Fu:2020lob}. The conformal time $\eta\gg\eta_{k}$ in Eq. \eqref{OmegaGW0} should be chosen earlier than the moment of matter-radiation equality, and of course late enough so that $\Omega_{\rm GW}(k,\eta)$ can be converged to a constant value.

\begin{figure*}
\begin{center}
\scalebox{0.9}[0.9]{\includegraphics{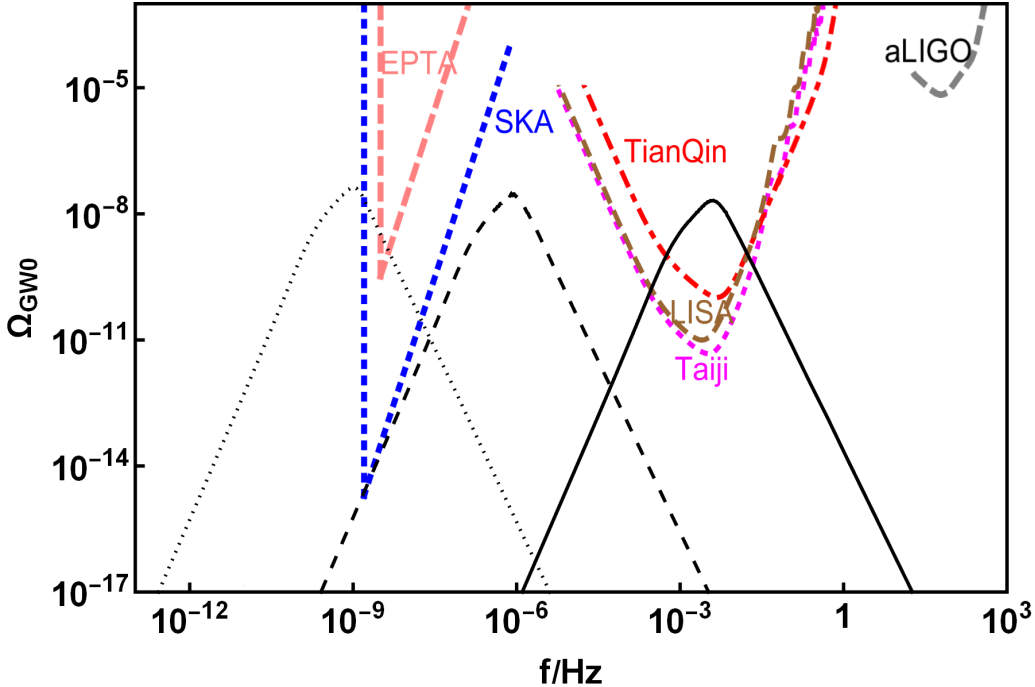}}
\caption{The present fractional energy density of the secondary GWs versus frequency. The solid, dashed, and dotted plots are corresponding to Case 1, Case 2, and Case 3, respectively.}
\label{fig:OmegaGW0}
\end{center}
\end{figure*}

In Fig. \ref{fig:OmegaGW0}, we plot the present fractional energy density of the secondary GWs for the three cases of our $\alpha$-attractor G-inflation model by using Eq. \eqref{OmegaGW0}. In the figure, we also have marginalized the sensitivity regions of the GWs detectors including European PTA (EPTA) \cite{Ferdman:2010xq, Hobbs:2009yy, McLaughlin:2013ira}, the Square Kilometer Array (SKA) \cite{Moore:2014lga}, Advanced Laser Interferometer Gravitational Wave Observatory (aLIGO) \cite{Harry:2010zz, TheLIGOScientific:2014jea}, Laser Interferometer Space Antenna (LISA) \cite{Danzmann:1997hm,Audley:2017drz}, TaiJi \cite{Hu:2017mde}, and TianQin \cite{Luo:2015ght}. We see in the figure that for all the three cases, the peak amplitude of the spectra is of order $10^{-8}$, but the peaks appear in different frequencies. In the Case 1, the peak takes place at the critical frequency $f_{c}\sim10^{-3}\,\mathrm{Hz}$, and so the result can be located inside the sensitivity regions of the space-based detectors LISA, TaiJi, and TianQin. For the Case 2, the peak takes place at $f_{c}\sim10^{-7}\,\mathrm{Hz}$, and the spectrum cannot be located within the joint region of anyone of the mentioned GWs detectors. The peak of Case 3 appears at the frequency $f_{c}\sim10^{-10}\,\mathrm{Hz}$, and the spectrum of this case can lie within the sensitivity region of EPTA and SKA. The exact values of the critical frequencies and peak heights for the three cases are tabulated in Table \ref{table:GWs}. Since some of these predictions can be located inside the sensitivity marginalized joints regions of some designed GWs detectors, it may be possible in future to check the consistency of our model in front of the observational data.

Another important observational criterion for the secondary GWs is the tilt of the spectrum of the present energy density fraction at different frequencies which may be appraised in light of the data from GWs detectors. Indeed, the recent studies imply that the power spectrum of $\Omega_{\rm GW_0}$ can be parameterized in terms of frequency as the power-law form $\Omega_{\rm GW_0}\sim f^{n}$, where $n$ is constant \cite{Xu:2019bdp, Fu:2019vqc, Kuroyanagi:2018csn}. In our work, we estimated the power $n$ in the ranges $f \ll f_c$, $f < f_c$, and $f > f_c$, which are denoted by $n_{f\ll f_{c}}$, $n_{f<f_{c}}$, and $n_{f>f_{c}}$, respectively. The numerical values of these parameters are presented in Table \ref{table:GWs}. The results obtained in the infrared regime $f \ll f_c$ satisfy appropriately the analytical expression $\Omega_{\rm GW_0}\sim f^{3-2/\ln\left(f_{c}/f\right)}$ obtained by \cite{Yuan:2019wwo, Cai:2019cdl}.

\begin{table}[ht!]
  \centering
  \caption{The frequencies and heights of the peak of the spectrum of the present fractional energy density of the secondary GWs for the three cases. The values of the tilt of the secondary GWs spectrum are also presented for the frequency ranges $f \ll f_c$, $f < f_c$, and $f > f_c$.}
\scalebox{1}[1] {
    \begin{tabular}{cccccc}
\hline
\hline
$\qquad$ \# $\qquad$ & $ \qquad f_{c} \qquad$ & $\qquad \Omega_{\rm GW_0}\left(f_{c}\right) \qquad$ & $\quad n_{f\ll f_{c}} \quad$ & $\quad n_{f<f_{c}} \quad$ & $\quad n_{f>f_{c}} \quad$\tabularnewline
\hline
Case 1 & $4.077\times10^{-3}$ & $2.093\times10^{-8}$ & $3.013$ & $1.505$ & $-2.647$\tabularnewline
\hline
Case 2 & $8.006\times10^{-7}$ & $3.012\times10^{-8}$ & $3.011$ & $1.512$ & $-2.724$\tabularnewline
\hline
Case 3 & $1.016\times10^{-10}$ & $4.284\times10^{-8}$ & $3.032$ & $1.477$ & $-2.753$\tabularnewline
\hline
\end{tabular}
    }
  \label{table:GWs}
\end{table}


\section{Conclusions}
\label{sec:con}

The PBHs formation due to the gravitational collapse of the inflationary curvature perturbations has been extensively studied over the decades. After the first GW discovery by the LIGO-Virgo Collaboration, interest in PBHs has renewed and the scenario of PBHs as DM candidate has attracted more and more attentions. In this work, we examined the generation of PBHs in the framework of $\alpha$-attractor G-inflation scenario. We considered the simplest form of the Galileon field Lagrangian described by Eq. (\ref{Lagrangian}) where $G_3(\phi,X)$ and $K(\phi,X)$ are functions in terms of Galileon field $\phi$ and the kinetic term $X\equiv -\frac{1}{2}g^{\mu\nu}\partial_\mu{\phi}\partial_\nu{\phi}$. Assuming that $G_3$ only depends on the scalar field $\phi$, the action (\ref{action0}) is transformed to Eq. (\ref{action}) which describes a noncanonical inflationary model with the kinetic term $\big(1-2G(\phi)\big)X$ in which we defined $G(\phi)\equiv dG_{3}/d\phi= G_{3,\phi}$. We took $G(\phi)$ as $G_I(\phi)\big(1+G_{II}(\phi)\big)$ where the function $G_I(\phi)$ and $G_{II}(\phi)$ are given in Eqs. (\ref{GI}) and (\ref{GII}). The quantity $G_I(\phi)$ is the base Galileon term and $G_{II}(\phi)$ is a function with a peak at $\phi=\phi_c$ which is almost negligible for the field values away from $\phi_c$. Therefore, for the field values away from $\phi_c$, the kinetic term $\big(1-2G(\phi)\big)X$ in Eq. (\ref{action}) nearly reduces to $\Big(2\alpha X/(1-\phi^{2}/3)^2\Big)$ which is the same as one in the $\alpha$-attractors scenario. We choose the potential in the form of $V(\phi)=V_{0}\phi^{2n}$ which is the simplest form of the potential in the $\alpha$-attractor models. By redefining $\phi/\sqrt{3}=1-e^{-\sqrt{\frac{2}{3\alpha}}\varphi}$, the kinetic term $\Big(2\alpha X/(1-\phi^{2}/3)^2\Big)$ reduces to the canonical form $-g^{\mu\nu}\partial_{\mu}\varphi\partial_{\mu}\varphi/2$. This class of $\alpha$-attractors is called $E$-model. For $n=1$ and $\alpha=1$ by fine-tuning of the parameters $\phi_c$, $\omega$, and $\sigma$, we found three successful parameter sets to form PBHs (see Tables \ref{Table1} and \ref{Table2}). We studied the inflationary dynamics for these three parameter sets by solving the field equations numerically. The evolution of the scalar field $\phi$, the slow-roll parameters $\varepsilon$ and $\delta_\phi$ as functions of the $e$-fold number $N$ are plotted in Figs. \ref{fig:phiN} and \ref{fig:SRparameters}. From Fig. \ref{fig:SRparameters}, we see that in the ultra slow-roll phase, the slow-roll condition $|\delta_\phi|<1$ is violated. Consequently, using Eq. (\ref{Ps1}) or Eq. (\ref{PsSR}) to find the power spectrum leads to wrong results. Thus, we computed the exact power spectrum by solving the Mukhanov-Sasaki equation numerically. The results are shown in Fig. \ref{psk} for the three parameter sets 1, 2, and 3. This figure indicates that on the large scales, the obtained power spectra are consistent with the CMB observation and on the small scales, when the inflaton experiences an ultra slow-roll phase, they enhance to the order of ${\cal O}(10^{-2})$ which is sufficiently large to provide PBHs production.

We further estimated the inflationary observables $n_s$, $r$, and $dn_s/d\ln k$ predicted by our model for the three cases 1, 2, and 3. Our results showed that the values of $r$, and $dn_s/d\ln k$ for parameter sets 1 and 2, are in agreement with the $68\%$ CL constraints of Planck 2018 TT+lowE data \cite{akrami2020planck}, while the values of $n_s$  satisfy the $95\%$ CL constraints of these data. For Case 3, the values of $n_s$ and $r$ are in well agreement with the $68\%$ CL constraints of Planck 2018 TT+lowE data \cite{akrami2020planck}, and the value of $dn_s/d\ln k$ is consistent with the 95\% CL constraint of these observational data (see Table \ref{Table2}).

With the help of Press-Schechter formalism and using the exact power spectra, we found the PBHs abundances for the three parameter sets 1, 2 and 3. The predicted PBHs abundances are plotted in Fig. \ref{fPBH-M}. The obtained numerical results are also shown in Table \ref{Table2}. Our analysis showed that the parameter set 1 produces PBHs with mass $M\simeq 8.06\times10^{-13} M_\odot$ which can explain around $98\%$ of the current DM content. Therefore, the produced PBHs in this class can be considered as a suitable candidate for DM. The formed PBHs in parameter set 2 have the mass of $ 1.78\times10^{-5}M_\odot$ which can explain the ultrashort-timescale microlensing events in OGLE data. In parameter set 3, we found PBHs with mass $M\simeq 12.99 M_\odot$. Fig. \ref{fPBH-M} indicates the peak of $f_{\rm{PBH}}$ does not exceed the upper limit on the merger rate obtained from the LIGO.

Moreover, we studied the induction of the secondary GWs accompanied by the PBHs formation in our $\alpha$-attractor G-inflation setup, and in particular we computed the present fractional energy density ($\Omega_{\rm GW_0}$) for the three parameter sets of our model. The spectrum of $\Omega_{\rm GW_0}$ exhibits a peak in its shape, and the peaks height for all the three cases is of order $10^{-8}$, but their frequencies are different. The frequencies of the peaks for Cases 1, 2, and 3 are $2.953\times10^{-3}Hz$, $8.017\times10^{-7}Hz$, and $5.848\times10^{-10}Hz$, respectively. The spectrum of $\Omega_{\rm GW_0}$ for Case 1 can be placed within the sensitivity region of LISA, TaiJi, and TianQin, and for Case 3 within the sensitivity regions of EPTA and SKA, while for Case 2, the spectrum is located completely outside of the sensitivity curves. Since the predictions of our $\alpha$-attractor G-inflation model can lie inside the sensitivity regions of some GWs detectors, therefore the viability of our model can be tested in light of the forthcoming observational data. We also estimated the tilt of the spectrum of secondary GWs in our setting for different regions of the frequency band. Our findings confirm that the power spectrum of $\Omega_{\rm GW_0}$ can be parameterized in terms of frequency as the power-law function $\Omega_{\rm GW_0}\sim f^{n}$. We calculated the values of the constant $n$ for different frequency bands for each case of our model, and showed that the results in the infrared regime $f \ll f_c$ satisfy properly the analytical expression $\Omega_{\rm GW_0}\sim f^{3-2/\ln\left(f_{c}/f\right)}$ presented by \cite{Yuan:2019wwo, Cai:2019cdl}.

Finally, it is important to note that we have assumed a Gaussian statistics of primordial scalar perturbations in our work. Since any non-Gaussianity can affect the PBHs abundance, it will be interesting to study the role of the non-Gaussianity on the number of PBHs and the induced GWs. We left this issue for future works.

\subsection*{Acknowledgements}
The authors thank the referee for his/her valuable comments.


%

\end{document}